\documentclass[twocolumn,twoside,prl,aps]{revtex4}
\usepackage{amsmath,mathrsfs,amsbsy,color,graphicx,bm,amsthm,amsfonts}
\usepackage{units}
\usepackage{bbm}

\begin{document}

\title{Single-photon quantum router with multiple output ports}
\author{Wei-Bin Yan}
\author{Heng Fan}
\email{hfan@iphy.ac.cn}
\affiliation{Beijing National Laboratory for Condensed Matter Physics, Institute of
Physics, Chinese Academy of Sciences, Beijing 100190, China}

\begin{abstract}
We study the multi-channel quantum routing of the single photons in a
waveguide-emitter system. The channels are composed by the waveguides and
are connected by intermediate two-level emitters. By adjusting the
intermediate emitters, the output channels of the input single photons can
be controlled. This is demonstrated for the cases of one output channel, two
output channels and the generic $N$ output channels. The results show that
the multi-channel quantum routing of single photons can be well achieved in
present system. This sheds light on the experimental realization of quantum
routing of single photons.
\end{abstract}

\maketitle

\emph{Introduction.}---Quantum routing of information from one sender to a
lot of receivers plays an essential role in the quantum network. Single
photons are suitable candidates for the carrier of quantum information due
to the fact that they propagate fast and interact rarely with the
environment. The manipulation of photons needs the field-matter interaction.
The photon transport in a one dimensional (1D) waveguides have been studied
extensively both in theory\cite%
{Shen2005ol,Shen2007pra,Lan,Roy2010prb,Fan2012prl,Paolo,Zhengprl,
Liao,Witthaut2010njp,liqiong} and in experiments \cite{1,2,Akimov,Bajcs,
Babinec,Claudon,Bleuse,Laucht} because the strong
coupling of the waveguide-emitter can be achieved. In the waveguide-emitter
system, the waveguides act as the channels and the emitters as the nods of
the quantum network. Based on these advantages, the quantum routing of
photons in the waveguide-emitter system is promising. Recently, the authors
in Ref. \cite{zhour} proposed a novel two output channel quantum routing of
single photons in a waveguide system. They connected two 1D waveguides by an
intermediate three-level system. Consequently, the input single photons can
be redirected into either of the two output channels with a maximal
probability of unity and more than $\frac{1}{2}$, respectively. It is
interesting if the photon can be redirected into either of the output
channels with an extremely high probability. Moreover, a more than two
output channel quantum routing of the single photons will be of great
interest.

For these purposes, we propose a scheme to achieve the quantum routing of
single photons from one input channel into $N$ output channels. In our
scheme, the $i$th output channel is connected with the input channel by an
intermediate two-level system (TLS). We find the generic solution of the
probabilities of the photon in each channel in the long-time limit. For the
single output channel quantum routing, the quantum interferences redirect
the input photon into the output channel completely when the intermediate
TLS resonantly interacts with the two output channels with the same
strength. By adjusting the parameters, the input photon can be redirected
into the output channel with desired probabilities. The single output
channel routing properties can be modified when another additional TLS is
coupled to the input channel. In the two output channel case, the photon can
be redirected into any of the channels with an approximate unity
probability. The photon can also be redirected completely or not completely
into the two output channels, where the probabilities of the photon can be
tuned. In the generic $N$ output channel case, the quantum interferences
prevent the photon being redirected into the other channels except the input
channel when all the TLSs resonantly interact with the channels with equal
strengths for a large value of $N$. They also completely prevent the photon
being directed back into the input channel for suitable parameters for any
value of $N$. The photon can be redirected into a desired channel with an
approximate unity probability. Thus, the $N$ output channel quantum routing
of the single photons can be achieved in our scheme.

\begin{figure}[tbp]
\includegraphics*[width=7cm, height=6cm]{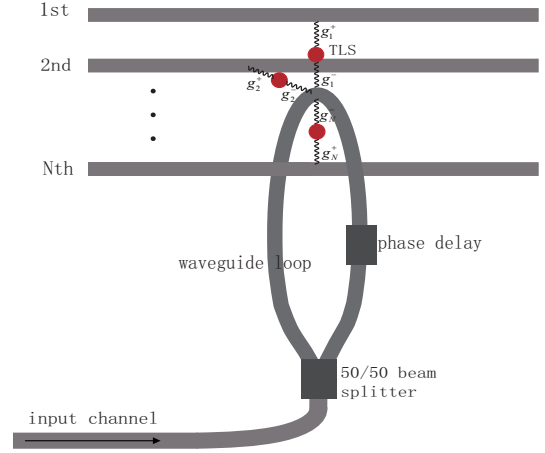}
\caption{Schematic diagram of the multi-channel quantum routing of the
single photons. $N$ 1D waveguides are connected with the input channel by $N$
intermediate two-level systems.}
\end{figure}

\emph{System description.}---The schematic diagram of the system under
consideration is shown in Fig. 1. The system consists of an input 1D
semi-infinite channel plus $N$ output 1D infinite waveguides. The
input channel is a sagnac interferometer \cite{sag1,sag2,sag3} composed by a
semi-infinite waveguide,
a 50:50 beam splitter and a waveguide loop. The input channel connects the $i
$th infinite waveguide by an intermediate TLS at the middle point of the
waveguide loop. For simplicity, we label the interaction position $x=0$. The
TLS can be a manual atom-like object or a cavity-atom dressed system. In the
single-excitation case, the TLS can also be a cavity. There are two
continuum of modes, right-moving modes and left-moving modes, in the 1D
waveguide. The clockwise and counter clockwise-moving modes in the waveguide
loop can be considered as the left- and right-moving modes. It is convenient
to bring in the even and odd operators as $a_{ek}=\frac{1}{\sqrt{2}}%
(a_{rk}+a_{lk})$ and $a_{ok}=\frac{1}{\sqrt{2}}(a_{rk}-a_{lk})$, with $a_{rk}
$ ($a_{lk}$) being the annihilation operator for the right (left)-moving
mode with the frequency $v_{g}k$ in the waveguide. Hereafter we will take
the photon group velocity $v_{g}=1$. In the even and odd picture, the TLSs
only interact with the even mode, while the odd mode only contributes to the
free energy part in the Hamiltonian. Therefore, it is enough to study the
dynamics of the even mode. The even part of the Hamiltonian can be written
as ($\hbar =1$)%
\begin{eqnarray}
H &=&\int_{-\infty }^{\infty }dkka_{k}^{\dagger
}a_{k}+\sum_{i}[\int_{-\infty }^{\infty }dkka_{i,k}^{\dagger }a_{i,k}+\omega
_{i}\sigma _{i}^{+}\sigma _{i}^{-}  \notag \\
&&+\int_{-\infty }^{\infty }dk(g_{i}^{+}a_{i,k}+g_{i}^{-}a_{k})\sigma
_{i}^{+}+h.c.]\text{,}
\end{eqnarray}%
with $a_{k}^{\dagger }$ ($a_{i,k}^{\dagger }$) being the even mode creation
operator in the input ($i$th output) waveguide, and $\omega _{i}$ being the $%
i$th TLS transition frequency. We have taken the energies of the TLS ground
states zero. The terms of the second line in Hamiltonian (1) represent the
interaction of the TLSs with the waveguides. $\frac{g_{i}^{-}}{\sqrt{2}}$
and $\frac{g_{i}^{+}}{\sqrt{2}}$ are coupling strengths of the $i$th TLS to
the input waveguide and the $i$th output waveguide, respectively. The
coupling strengths can be written as $g_{i}^{\zeta }=\sqrt{\frac{\gamma
_{i}^{\zeta }}{2\pi }}(\zeta =\pm )$, with $\gamma _{i}^{\zeta }$ being the
decay rate from the $i$th TLS to the waveguides. We have made two
approximations: one is extending the frequency integration to $\pm \infty $,
the other is the coupling strengths independent of the frequency, which is
equivalent to the Markovian approximation. These approximations are valid
since we will focus on the pulse with a narrow frequency width around the
carrier frequency.

The arbitrary state of the system in the single-excitation subspace has the
form of $\left| \Psi (t)\right\rangle =\int dk\alpha _{k}(t)a_{k}^{\dagger
}+\sum_{i}[\int dk\alpha _{i,k}(t)a_{i,k}^{\dagger }+\beta _{i}(t)\sigma
_{i}^{+}]\left| \phi \right\rangle $, with $\alpha _{k}(t)$, $\alpha
_{i,k}(t)$ and $\beta _{i}(t)$ being the probability amplitudes. The state $%
\left| \phi \right\rangle $ represents that all the waveguides are in the
vacuum states and all the TLSs are in the ground states. If we inject a
photon into the input waveguide, an even mode quasi particle can be produced
at the middle point of the waveguide loop when the phases of the clockwise-
and counter clockwise-moving photons are equal. We assume that, initially, a
photon prepared in a wave packet with a Lorenzian spectrum is injected into
the input waveguide, while the TLSs and the output waveguides contain no
excitation, i. e. $\alpha _{k}(0)=\frac{\sqrt{\epsilon /\pi }}{k-\varpi
+i\epsilon }$,$\ \alpha _{i,k}(0)=0$, $\beta _{i}(0)=0$. $\epsilon $ and $%
\varpi $ are the spectral width and the center frequency of the input wave
packet, respectively. $\epsilon \rightarrow 0$ is the monochromatic limit.
The Shr\"{o}dinger equation gives a set of differential equations of
probability amplitudes. By performing the Laplace\ and inverse Laplace
transformation, the probability amplitudes can be found under the initial
condition. The details of the technique of Laplace transformation in a 1D
waveguide coupled to an emitter can be seen in Ref. \cite{Liao}. In the
long-time limit, we can obtain the probability amplitudes of the photon in
each channel as $\underset{\lim t\rightarrow \infty }{\alpha _{i,k}(t)}%
=\alpha _{i,k}\alpha _{k}(0)e^{-ikt}$ and $\underset{\lim t\rightarrow
\infty }{\alpha _{k}(t)}=\alpha _{k}\alpha _{k}(0)e^{-ikt}$ when $\underset{%
\lim t\rightarrow \infty }{\beta _{i}(t)}=0$ \cite{method}, with%
\begin{eqnarray}
\alpha _{i,k} &=&\frac{-\sqrt{\gamma _{i}^{-}\gamma _{i}^{+}}%
\prod\limits_{j\neq i}(i\delta _{j}+\frac{\gamma _{j}^{+}}{2})}{%
\prod\limits_{j=1..N}(i\delta _{j}+\frac{\gamma _{j}^{+}}{2}%
)+\sum\limits_{j=0..N}[\frac{\gamma _{j}^{-}}{2}\prod\limits_{j^{\prime
}\neq j}(i\delta _{j^{\prime }}+\frac{\gamma _{j^{\prime }}^{+}}{2})]}\text{,%
}  \notag \\
\alpha _{k} &=&\frac{\prod\limits_{j=1..N}(i\delta _{j}+\frac{\gamma _{j}^{+}%
}{2})-\sum_{j}[\frac{\gamma _{j}^{-}}{2}\prod\limits_{j^{\prime }\neq
j}(i\delta _{j^{\prime }}+\frac{\gamma _{j^{\prime }}^{+}}{2})]}{%
\prod\limits_{j=1..N}(i\delta _{j}+\frac{\gamma _{j}^{+}}{2})+\sum_{j}[\frac{%
\gamma _{j}^{-}}{2}\prod\limits_{j^{\prime }\neq j}(i\delta _{j^{\prime }}+%
\frac{\gamma _{j^{\prime }}^{+}}{2})]}\text{.}
\end{eqnarray}%
The detuning $\delta _{j}=\omega _{j}-k$. The output pulse has the same
shape with the input pulse. Here we have considered the conservation of
energy that the carrier frequencies of the input and output pulse are
equal.\ The injected photon will be redirected into the $i$th channel with
the probability $\left| \alpha _{i,k}\right| ^{2}$ and back into the input
channel with the probability $\left| \alpha _{k}\right| ^{2}$. The
probabilities relate to the detunings and coupling constants. In order to
study the quantum routing of single photon more clearly, it is necessary to
investigate the routing properties in Eqs. (2) for various numbers of the
output channel.

\emph{Single output channel.}---We start from the simplest case that the
input channel is connected with only one output channel by a TLS. The
intermediate TLS can absorb the input photon and then reemit it into the
input and output channels.The probability amplitudes are found as $\alpha
_{1,k}^{(1)}=\frac{-\sqrt{\gamma _{1}^{-}\gamma _{1}^{+}}}{i\delta _{1}+%
\frac{\gamma _{1}^{+}}{2}+\frac{\gamma _{1}^{-}}{2}}$, and $\alpha
_{k}^{(1)}=\frac{i\delta _{1}+\frac{\gamma _{1}^{+}}{2}-\frac{\gamma _{1}^{-}%
}{2}}{i\delta _{1}+\frac{\gamma _{1}^{+}}{2}+\frac{\gamma _{1}^{-}}{2}}$. To
make a distinction between the amplitudes in different $N$ cases, we make a
superscript $n$ in $\alpha _{1,k}^{(n)}$ and $\alpha _{k}^{(n)}$ standing
for the $N=n$ case. Obviously, only when $\frac{\gamma _{1}^{+}}{\gamma
_{1}^{-}}=1$ and $\delta _{1}=0$, we can find $\left| \alpha
_{1,k}^{(1)}\right| ^{2}=1$ and $\left| \alpha _{k}\right| ^{2}=0$. That is
to say, the input photon will be redirected into the output channel
completely only when the input photon resonantly interacts with the TLS and
the TLS decays to the input and output channels at the same rate. This is
due to the quantum interferences. When the detunings are large enough, i. e.
$\delta _{1}\gg \{\frac{\gamma _{1}^{+}}{2},\frac{\gamma _{1}^{-}}{2}\}$,
the photons will be back into the input channel with an approximate unity
probability. Here we bring in a parameter $\eta _{i}=\frac{\max \{\gamma
_{i}^{+},\gamma _{i}^{-}\}}{\min \{\gamma _{i}^{+},\gamma _{i}^{-}\}}$ to
measure the difference between the coupling strengths of the $i$th TLS to
the input and the $i$th waveguides. When $\eta _{1}\gg 1$,\ the photons will
be back into the input channel with an approximate unity probability.
Although the decay rate of the TLS to the input channel is much smaller than
the decay rate to the other channel, the photon is prevented being
redirected into the other channel. $\eta _{1}\rightarrow \infty $\
corresponds to the limit that one of the two waveguides is decoupled to the
TLS. In this case, the input and output channels are not connected, and the
photon will be back into the input channel completely. To show the details
of the quantum routing of the single phonon in the $N=1$ case, we plot the
probability $\left| \alpha _{1,k}^{(1)}\right| ^{2}$ against the detuning
and decay rates in Fig. 2. The probability $\left| \alpha _{k}^{(1)}\right|
^{2}$\ is not plotted here because $\left| \alpha _{k}^{(1)}\right|
^{2}=1-\left| \alpha _{1,k}^{(1)}\right| ^{2}$.\ The probabilities of the
photon in the input and output channels can be controlled by adjusting the
detuning and coupling strengths. The single photon can be directed into
either of the two channels completely or directed into the two channels with
a desired probability. Hence, we can achieve the quantum routing of the
single photon when $N=1$.

\begin{figure}[tbp]
\includegraphics*[bb=5 80 591 413,width=8cm, height=5.5cm]{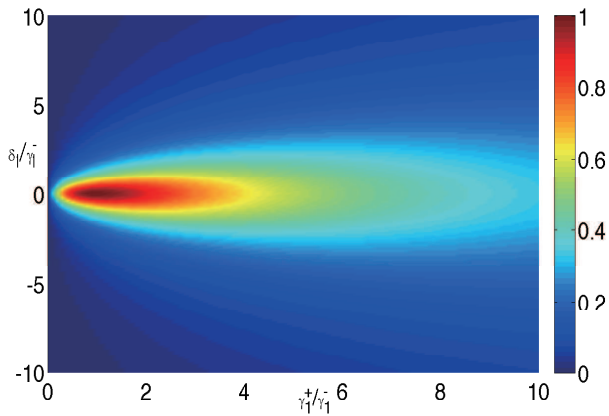}
\caption{Probability of the single photon in the 1st output channel in the
long-time limit $\left| \protect\alpha _{1,k}^{(1)}\right| ^{2}$ against the
detuning and decay rate when $N=1$.}
\end{figure}

The $N=1$ case is analogous to a on chiral waveguide coupled to a $\Lambda $%
-type three-level quantum emitter. The states $a_{k}^{\dagger }\left| \phi
\right\rangle $, $a_{1,k}^{\dagger }\left| \phi \right\rangle $, $\sigma
_{1}^{+}\left| \phi \right\rangle $ are mapped to the states of the $\Lambda
$-type emitter-waveguide system well. In Ref. \cite{Witthaut2010njp,fc}, the
authors investigated the single-photon transport in the $\Lambda $-type
emitter-waveguide system. The similar physics can be seen in the two schemes.

\emph{Two output channels.}---When the input channel is connected with two
output channels by two TLSs, the probability amplitudes are obtained as $%
\alpha _{1,k}^{(2)}=\frac{-\sqrt{\gamma _{1}^{-}\gamma _{1}^{+}}(i\delta
_{2}+\frac{\gamma _{2}^{+}}{2})}{(i\delta _{1}+\frac{\gamma _{1}^{+}}{2}+%
\frac{\gamma _{1}^{-}}{2})(i\delta _{2}+\frac{\gamma _{2}^{+}}{2}+\frac{%
\gamma _{2}^{-}}{2})-\frac{\gamma _{1}^{-}\gamma _{2}^{-}}{4}}$, and $\alpha
_{k}^{(2)}=\frac{(i\delta _{1}+\frac{\gamma _{1}^{+}}{2}-\frac{\gamma
_{1}^{-}}{2})(i\delta _{2}+\frac{\gamma _{2}^{+}}{2}-\frac{\gamma _{2}^{-}}{2%
})-\frac{\gamma _{1}^{-}\gamma _{2}^{-}}{4}}{(i\delta _{1}+\frac{\gamma
_{1}^{+}}{2}+\frac{\gamma _{1}^{-}}{2})(i\delta _{2}+\frac{\gamma _{2}^{+}}{2%
}+\frac{\gamma _{2}^{-}}{2})-\frac{\gamma _{1}^{-}\gamma _{2}^{-}}{4}}$.
Because the expression of $\alpha _{1,k}^{(2)}$ is symmetrical to $\alpha
_{2,k}^{(2)}$, the study of the properties $\left| \alpha
_{1,k}^{(2)}\right| ^{2}$ and $\left| \alpha _{k}^{(2)}\right| ^{2}$\ is
enough. The large detuning between the $1$st TLS and the input photon
prevents the photon being redirected into the $1$st channel. Especially,
when $i\delta _{2}+\frac{\gamma _{2}^{+}}{2}-\frac{\gamma _{2}^{-}}{2}=0$,
it also prevents the photon being directed into the input channel. As a
result, for an extremely large value of $\delta _{1}$, the photon will be
redirected into the $2$nd channel with an approximate unity probability when
$i\delta _{2}+\frac{\gamma _{2}^{+}}{2}-\frac{\gamma _{2}^{-}}{2}=0$. And
when the detunings $\delta _{1}$ and $\delta _{2}$ are large enough, the
photon will be directed into the input channel with an approximate unity
probability. Besides, the photon distribution can be affected significantly
by the coupling strengths. For example, when $i\delta _{2}+\frac{\gamma
_{2}^{+}}{2}-\frac{\gamma _{2}^{-}}{2}=0$, the large value of $\gamma
_{1}^{+}$ prevents the photon being directed into both the input and $1$st
channels. When $\eta _{1}=\eta _{2}=1$ and $\delta _{1}=\delta _{2}=0$, the
input photon will be directed into the input channel with a small
probability and be redirected into the other two channels averagely with a
large probability. If we do not want the input photon to be back into the
input channel, it is easy to choose the appropriate parameters which satisfy
$\alpha _{k}^{(2)}=0$. For example, when $\delta _{1}=\delta _{2}$, the
relation $(\gamma _{1}^{+}-\gamma _{1}^{-})(\gamma _{2}^{+}-\gamma
_{2}^{-})=\gamma _{1}^{-}\gamma _{2}^{-}$ can be easily satisfied. In this
case, the photon distributions in the two output waveguides are different
for various values of the decay rates. Especially, when $\gamma
_{1}^{+}=2\gamma _{1}^{-}$ and $\gamma _{2}^{+}=2\gamma _{2}^{-}$, the input
single photon is redirected into the two output channels with equal
probability $\frac{1}{2}$. These results provide the two output channel
quantum routing of single photons. To see the details, we plot the
probabilities $\left| \alpha _{1,k}^{(2)}\right| ^{2}$ and $\left| \alpha
_{k}^{(2)}\right| ^{2}$ against the detunings and coupling strengths in Fig.
3. Fig. 3(a) and 3(b) are $\left| \alpha _{1,k}^{(2)}\right| ^{2}$ and $%
\left| \alpha _{k}^{(2)}\right| ^{2}$, respectively, against the detuning
and coupling strength of the 2nd TLS when $\delta _{1}=0\ $and $\gamma
_{1}^{+}=\gamma _{1}^{-}=\gamma _{2}^{-}$. The large values of $\delta _{2}$
and $\gamma _{2}^{+}$ have a constructive effect on redirecting the photon
into the 1st channel. However, when $\delta _{2}$ and $\gamma _{2}^{+}$ is
small enough, the photon will also be back into the input channel. This will
be studied in detail below. Fig. 3(c) and 3(d) are $\left| \alpha
_{1,k}^{(2)}\right| ^{2}$ and $\left| \alpha _{k}^{(2)}\right| ^{2}$,
respectively, against the detunings $\delta _{1}$ and $\delta _{2}$, when
the two TLSs are coupled to the waveguides with equal strengths. The two
output channel quantum routing can be achieved well according to the results
obtained above.

\begin{figure}[tbp]
\includegraphics*[bb=14 60 594 374,width=4.1cm, height=3.2cm]{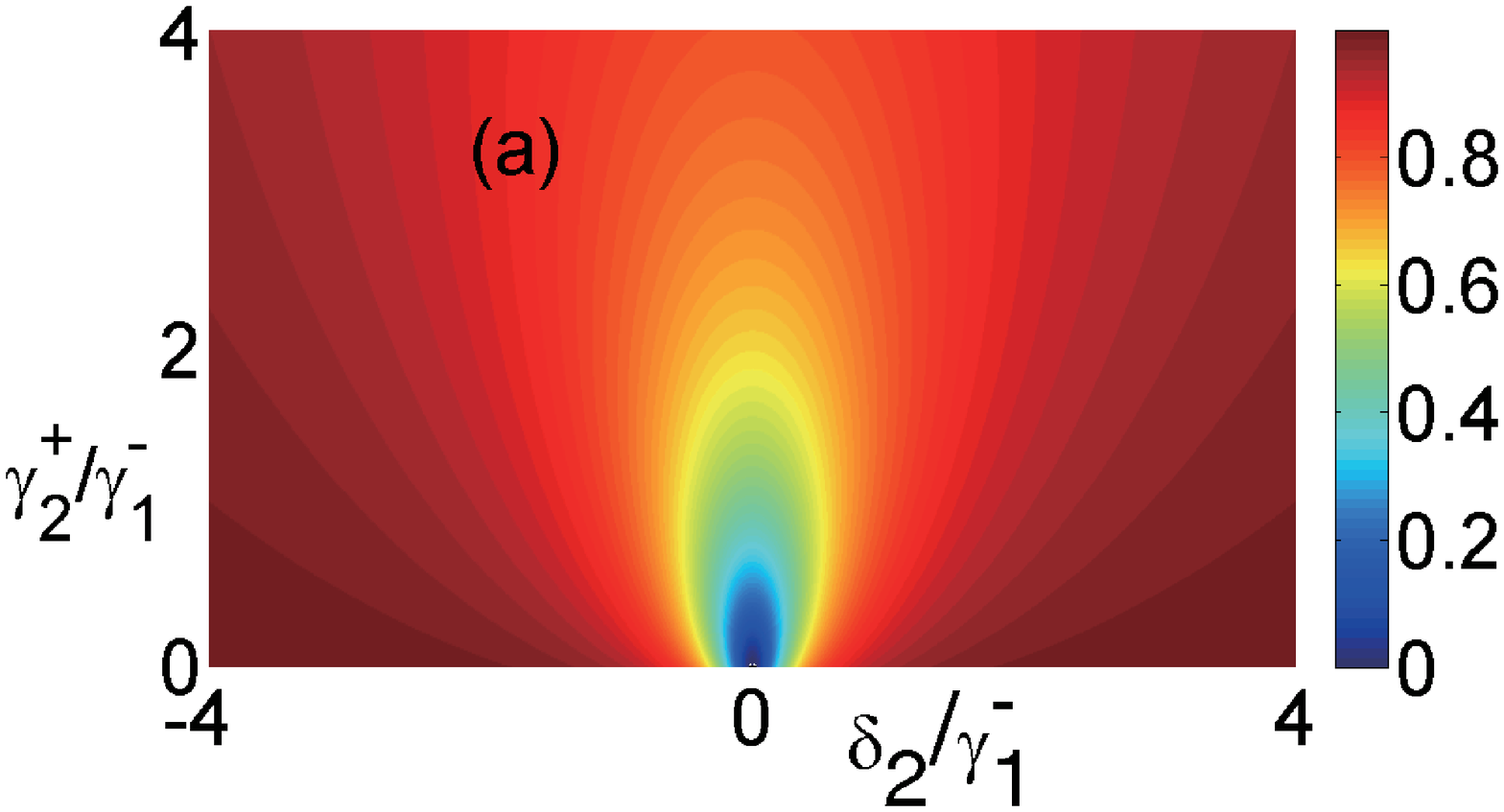} %
\includegraphics*[bb=0 46 594 356,width=4.1cm, height=3.2cm]{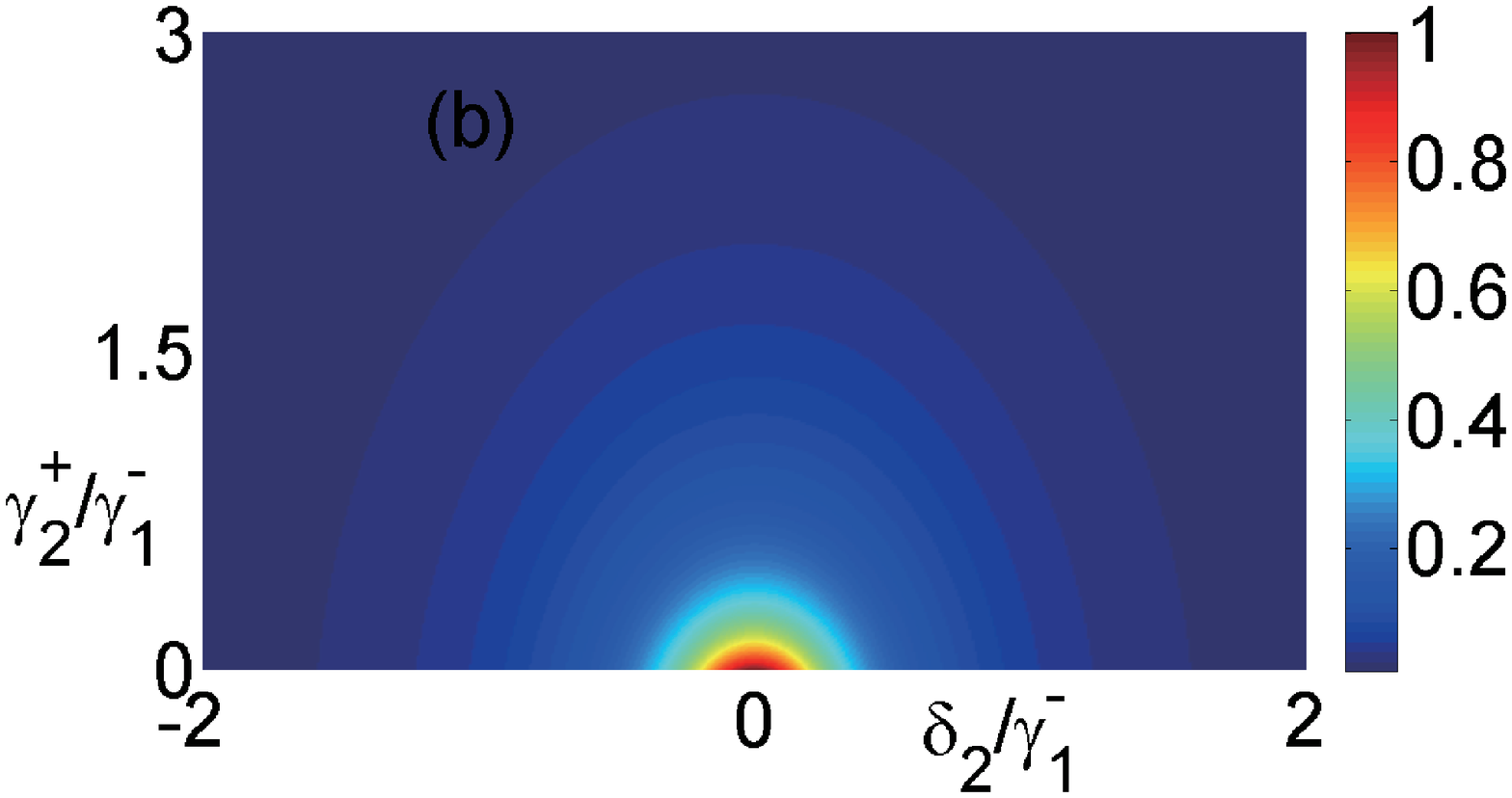} %
\includegraphics*[bb=4 71 595 408,width=4.1cm, height=3.2cm]{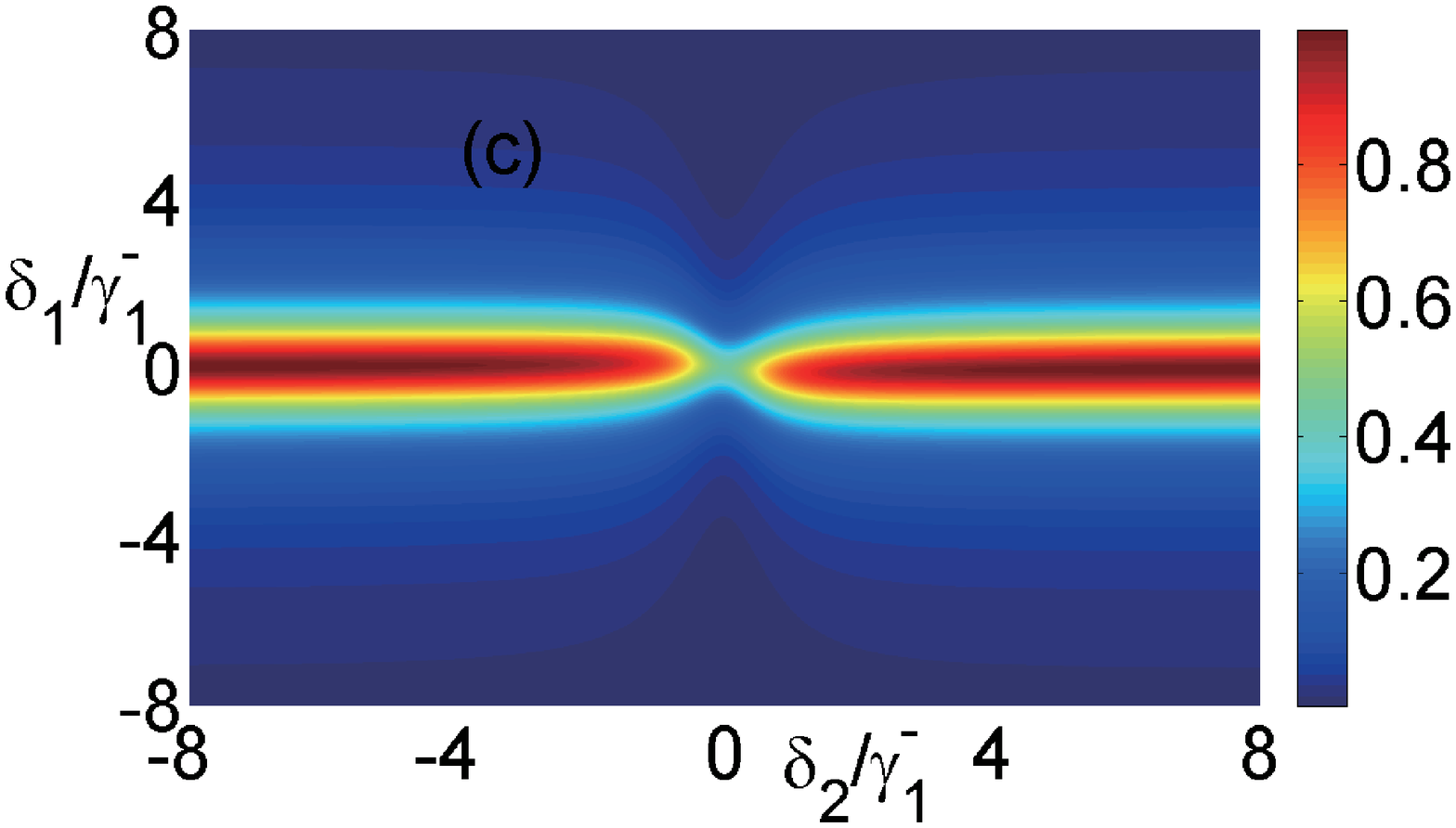} %
\includegraphics*[bb=8 85 592 416,width=4.1cm, height=3.2cm]{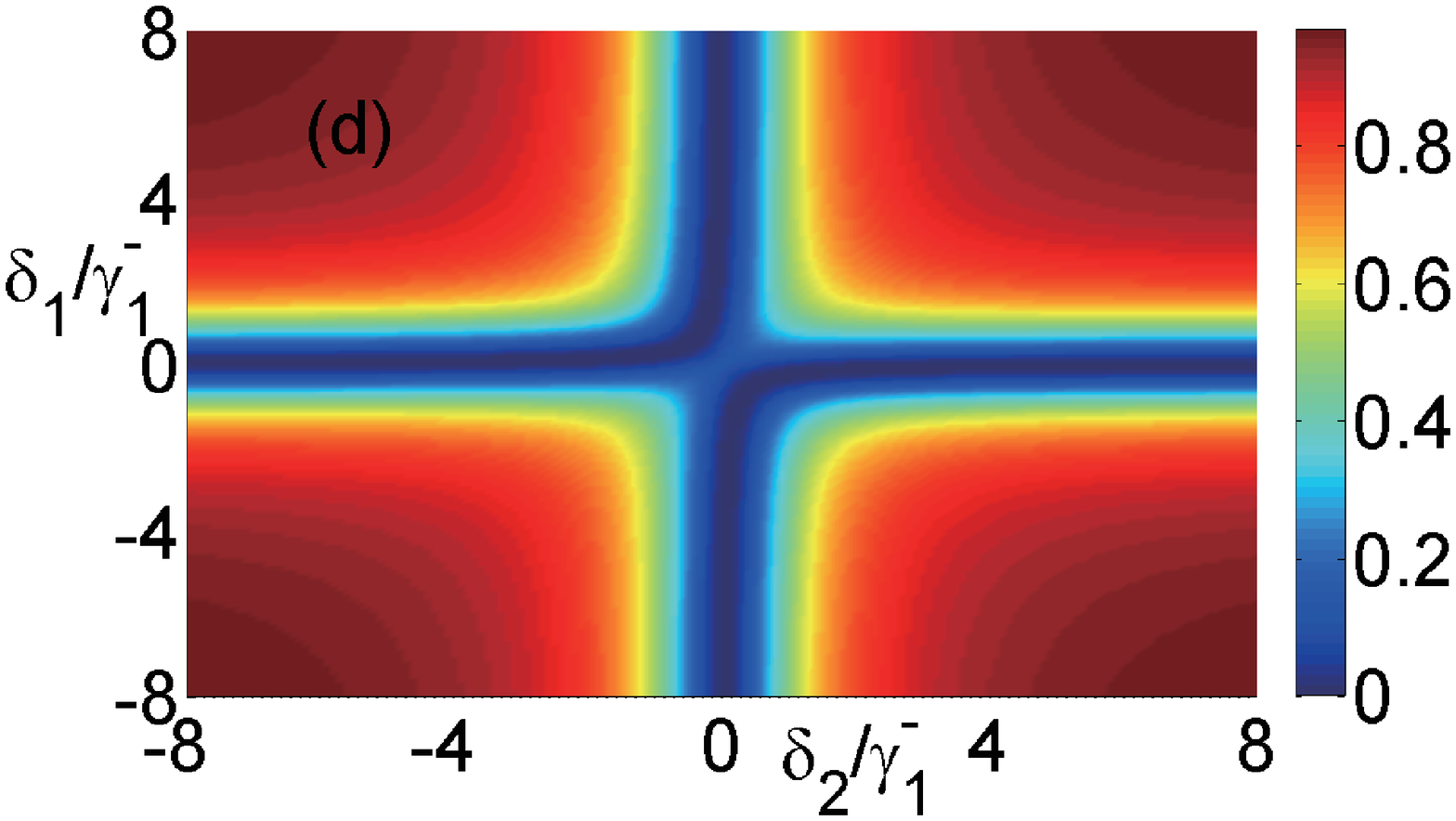} %
\caption{Probabilities $\left| \protect\alpha _{1,k}^{(2)}\right| ^{2}$ and $%
\left| \protect\alpha _{k}^{(2)}\right| ^{2}$ against the detunings and
decay rates when $N=2$. (a) and (c) denote the probability $\left| \protect%
\alpha _{1,k}^{(2)}\right| ^{2}$ against the parameters, and (b) and (d)
denote $\left| \protect\alpha _{k}^{(2)}\right| ^{2}$. (a) and (b) are the
probabilities against $\protect\delta _{2}$ and $\protect\gamma _{2}^{+}$
when $\protect\delta _{1}=0$ and $\protect\gamma _{1}^{-}=\protect\gamma %
_{1}^{+}=\protect\gamma _{2}^{-}$. (c) and (d) are the probabilities against
$\protect\delta _{1}$ and $\protect\delta _{2}$ when $\protect\gamma %
_{1}^{-}=\protect\gamma _{1}^{+}=\protect\gamma _{2}^{-}=\protect\gamma %
_{2}^{+}$.}
\end{figure}

It is necessary to study a special case of $N=2$, that is, the $2$nd TLS is
decoupled to the $2$nd output channel, i.e. $\gamma _{2}^{+}=0$. This can be
understood that an additional TLS is coupled to the input channel in the$\
N=1$ case. The additional TLS will modify the system behavior. For example,
when $\eta _{1}=1$ and $\delta _{1}=0$, we find $\alpha _{1,k}^{(2)}=\frac{%
-i4\delta _{2}\gamma _{1}^{-}}{4i\delta _{2}\gamma _{1}^{-}+\gamma
_{1}^{-}\gamma _{2}^{-}}$ and $\alpha _{k}^{(2)}=\frac{-\gamma
_{1}^{-}\gamma _{2}^{-}}{4i\delta _{2}\gamma _{1}^{-}+\gamma _{1}^{-}\gamma
_{2}^{-}}$. Hence, when the additional TLS resonantly interacts with the
input photon, i. e. $\delta _{2}=0$, the photon is directed into the input
channel compared with the $N=1$ case, in which the photon is redirected into
the 1st output channel. This can be seen in Fig. 3(a) and 3(b). When $\delta
_{2}$ is large enough, the input photon will be almost completely
redirected, mapped to the $N=1$ case. By adjusting the additional TLS, the
photon distribution in the input and output channels can be controlled. This
provides more control to the single output channel quantum routing of the
single photon.

\emph{N output channels.}---We proceed to study the general case that the
input channel connects with $N$ output channels by $N$ TLSs. Let's first
consider the simplest case that all of the $N$ extra channels are identical.
That is to say, all the TLS are identical, all the decay rates to the input
channel are identical, and all the decay rates to the output channels are
identical. We label $\delta _{i}=\delta $, $\gamma _{i}^{+}=\gamma ^{+}$,
and $\gamma _{i}^{-}=\gamma ^{-}$. The probability amplitudes are obtained
as $\alpha _{i,k}^{(N)}=\frac{-2\sqrt{\gamma ^{-}\gamma ^{+}}}{2i\delta
+\gamma ^{+}+N\gamma ^{-}}$, and $\alpha _{k}^{(N)}=\frac{2i\delta +\gamma
^{+}-N\gamma ^{-}}{2i\delta +\gamma ^{+}+N\gamma ^{-}}$. It is noted that
when all the TLSs interact resonantly with the input photon and the coupling
strengths satisfy $\gamma ^{+}=N\gamma ^{-}$, the interferences prevent the
input photon being directed into the input channel and redirect the input
photon into each of the $N$ output channels with equal probability $\frac{1}{%
N}$. When all the coupling strengths are equal to each other and all the
detunings are zero, the probabilities are obtained as $\left| \alpha
_{i,k}^{(N)}\right| ^{2}=(\frac{2}{1+N})^{2}$ and $\left| \alpha
_{k}^{(N)}\right| ^{2}=(\frac{1-N}{1+N})^{2}$. As discussed above, the
quantum interferences redirect the input photon completely from the input
channel into the other channel when $N=1$. As the number of the output
channels increases, the probability of the photon back into the input
channel increases. When $N=3$, the input photon is distributed in each of
the four channels, including the input and output channels, with equal
probability $\frac{1}{4}$. When the number of the output channels is large
enough, the quantum interferences direct the input photon back into the
input channel almost completely.

We now study a simple case to illustrate how to redirect the input photon
into the desired channel. For convenience, we assume that $N-1$ of the $N$
channels are identical except the $m$th channel. we label $\delta _{i\neq
m}=\delta ^{\prime }$, $\gamma _{i\neq m}^{+}=\gamma ^{+\prime }$, and $%
\gamma _{i\neq m}^{-}=\gamma ^{-\prime }$. The probability amplitudes are
obtained as $\alpha _{m,k}^{(N)}=\frac{-\sqrt{\gamma _{m}^{-}\gamma _{m}^{+}}%
(i\delta ^{\prime }+\frac{\gamma ^{+\prime }}{2})}{(i\delta _{m}+\frac{%
\gamma _{m}^{+}}{2})(i\delta ^{\prime }+\frac{\gamma ^{+\prime }}{2}+\frac{%
(N-1)\gamma ^{-\prime }}{2})+\frac{\gamma _{m}^{-}}{2}(i\delta ^{\prime }+%
\frac{\gamma ^{+\prime }}{2})}$ and $\alpha _{k}^{(N)}=\frac{(i\delta _{m}+%
\frac{\gamma _{m}^{+}}{2})(i\delta ^{\prime }+\frac{\gamma ^{+\prime }}{2}-%
\frac{(N-1)\gamma ^{-\prime }}{2})-\frac{\gamma _{m}^{-}}{2}(i\delta
^{\prime }+\frac{\gamma ^{+\prime }}{2})}{(i\delta _{m}+\frac{\gamma _{m}^{+}%
}{2})(i\delta ^{\prime }+\frac{\gamma ^{+\prime }}{2}+\frac{(N-1)\gamma
^{-\prime }}{2})+\frac{\gamma _{m}^{-}}{2}(i\delta ^{\prime }+\frac{\gamma
^{+\prime }}{2})}$. When the decay rates $\gamma _{m}^{-}=\gamma _{m}^{+}$
are much larger than other parameters, $\alpha _{m,k}^{(N)}\approx -\frac{%
i\delta ^{\prime }+\frac{\gamma ^{+\prime }}{2}}{i\delta ^{\prime }+\frac{%
\gamma ^{+\prime }}{2}+\frac{(N-1)\gamma ^{-\prime }}{4}}$. The photon will
be redirected into the $m$th channel almost completely when $2\gamma
^{+\prime }\gg (N-1)\gamma ^{-\prime }$. The limit of this condition is that
the $N-1$ TLSs are decoupled to the input channel. Besides, when all the
detunings are large enough, the photon will be directed into the input
channel with an approximate unity probability. When $\gamma _{m}^{-}=\gamma
_{m}^{+}$, $\delta _{m}=0$, and $\delta ^{\prime }$ is large enough, the
photon will be redirected into the $m$th channel with an approximate unity
probability. Similar to the $N=2$ case, the photon can be redirected
completely into the output channels where the photon probabilities can be
controlled. As a consequence, the $N$ output channel quantum routing of the
single photons can be achieved.

\emph{Conclusions.}---We propose a scheme for the quantum routing of the
single photons from one channel to many channels. The channels consisting of
waveguides are connected by intermediate TLSs. We consider that a single
photon is initially injected into the input channel. After interaction, the
TLSs can absorb the photon and reemit it into the channels. By solving the
Shr\"{o}dinger equation, the photon probability amplitudes in any of the
channels is obtained in the long-time limit. By studying the amplitudes, we
investigate the quantum routing of the single photons. We study the cases of
the single output channel, the two output channels, and the generic $N$
output channels in detail and show that the quantum routing can be achieved
in our scheme. We hope that our scheme will be realized in experiment.

\emph{Acknowledgement.}---This work is supported by ``973'' program
(2010CB922904), grants from Chinese Academy of Sciences, NSFC (11175248).

\emph{Methods.}---Here we give a brief description of the
calculation process. By performing the Laplace transformation, the equations
of the probability amplitudes under the initial condition can be written as%
\begin{eqnarray}
\alpha _{k}(s) &=&\frac{\sqrt{\pi /\epsilon }}{(s+i\omega _{k})(\omega
_{k}-\varpi +i\epsilon )}-\sum_{j}\frac{ig_{j}^{-}}{s+i\omega _{k}}\beta
_{j}(s)  \notag \\
\alpha _{i,k}(s) &=&\frac{-ig_{i}^{+}}{s+i\omega _{i,k}}\beta _{i}(s)  \notag
\\
0 &=&(s+i\omega _{i}+\frac{\gamma _{i}^{+}}{2})\beta _{i}(s)+\sum_{j}\frac{%
\sqrt{\gamma _{i}^{-}\gamma _{j}^{-}}}{2}\beta _{j}(s)  \notag \\
&&+\frac{2\pi g_{i}^{-}\sqrt{\pi /\epsilon }}{s+\epsilon +i\varpi }
\end{eqnarray}

We can find the solution of $\beta _{i}(s)$ from the last equation in Eqs.
(3) and then obtain $\alpha _{k}(s)$ and $\alpha _{i,k}(s)$. For simplicity,
we label $a_{i}=s+i\omega _{i}+\frac{\gamma _{i}^{+}}{2}$, $b_{i}=\sqrt{%
\frac{\gamma _{i}^{-}}{2}}$, and $c_{i}=\frac{2\pi g_{i}^{-}\sqrt{\pi
/\epsilon }}{s+\epsilon +i\varpi }$. Hence, the last equation can be written
as
\begin{equation}
a_{i}\beta _{i}(s)+b_{i}\sum_{j}b_{j}\beta _{j}(s)+c_{i}=0
\end{equation}%
From Eq. (4), we can find $\sum_{j}b_{j}\beta _{j}(s)=-\sum_{i}\frac{%
b_{i}^{2}}{a_{i}}\sum_{j}b_{j}\beta _{j}(s)-\sum_{i}\frac{c_{i}b_{i}}{a_{i}}$
to obtain the expression of $\sum_{j}b_{j}\beta _{j}$. By substituting the
solution of $\sum_{j}b_{j}\beta _{j}(s)$ into Eq. (4), we can obtain%
\begin{eqnarray*}
\beta _{i}(s) &=&-\frac{2\pi \sqrt{\pi /\epsilon }}{s+\epsilon +i\varpi } \\
&&\times \frac{g_{i}^{-}\prod_{j\neq i}(s+i\omega _{j}+\frac{\gamma _{j}^{+}%
}{2})}{\prod\limits_{j=1..N}(s+i\omega _{j}+\frac{\gamma _{j}^{+}}{2}%
)+\sum\limits_{j=1..N}[\frac{\gamma _{j}^{-}}{2}\prod\limits_{j^{\prime
}\neq j}(s+i\omega _{j^{\prime }}+\frac{\gamma _{j^{\prime }}^{+}}{2})]}%
\text{.}
\end{eqnarray*}%
Using the inverse Laplace transformation, the probability amplitudes $\beta
_{i}(t)$, $\alpha _{i,k}(t)$ and $\alpha _{k}(t)$ in the long-time limit can
be obtained. If $\beta _{i}(t)=0$ in the long-time limit, we can take the
conversation of energy condition $\omega _{i,k}=\omega _{k}$. In reality,
the TLSs are in their ground states in the long-time limit due to the
interaction with the continuum of modes. By the way, for the cases we
discussed in detail in the main text, the fact that $\underset{\lim
t\rightarrow \infty }{\beta _{i}(t)}=0$ can be easily verified from the
expression of $\beta _{i}(s)$.

\end{document}